# Conduction modulation of solution-processed two-dimensional materials

*Songwei Liu, Xiaoyue Fan, Yingyi Wen, Pengyu Liu, Yang Liu, Jingfang Pei, Wenchen Yang, Lekai Song, Danmei Pan, Teng Ma, Yue Lin, Gang Wang\*, and Guohua Hu\**

S. Liu, Y. Wen, P. Liu, Yang Liu, J. Pei, L. Song, G. Hu

Department of Electronic Engineering, The Chinese University of Hong Kong, Shatin, N. T., Hong Kong S. A. R., China

X. Fan, W. Yang, G. Wang

Centre for Quantum Physics, Key Laboratory of Advanced Optoelectronic Quantum Architecture and Measurement (MOE), School of Physics, Beijing Institute of Technology, Beijing 100081, China

Yang Liu

Shun Hing Institute of Advanced Engineering, The Chinese University of Hong Kong, Shatin, N. T., Hong Kong S. A. R., China

D. Pan, Yue Lin

CAS Key Laboratory of Design and Assembly of Functional Nanostructures, and State Key Laboratory of Structural Chemistry, Fujian Institute of Research on the Structure of Matter, Chinese Academy of Sciences, Fuzhou, Fujian 350002, China

T. Ma

Department of Applied Physics, Hong Kong Polytechnic University, Hung Hom, Kowloon, Hong Kong S. A. R., China

E-mail: ghhu@ee.cuhk.edu.hk, gw@bit.edu.cn

**Abstract**

Solution-processed two-dimensional (2D) materials hold promise for their scalable applications. However, the random, fragmented nature of the solution-processed nanoflakes and the poor percolative conduction through their discrete networks limit the performance of the enabled devices. To overcome the problem, we report conduction modulation of the solution-processed 2D materials via the Stark effect. Using liquid-phase exfoliated molybdenum disulfide ($MoS_2$) as an example, we demonstrate nonlinear conduction modulation with a switching ratio of $>10^5$ by the local fields from the interfacial ferroelectric P(VDF-TrFE). Through density-functional theory calculations and *in situ* Raman scattering and photoluminescence spectroscopic analysis, we understand the modulation arises from a charge redistribution in the solution-processed $MoS_2$. Beyond $MoS_2$, we show the modulation may be viable for the other solution-processed 2D materials and low-dimensional materials. The effective modulation can open their electronic device applications.

**1. Introduction**

Two-dimensional (2D) materials with a wide spectrum of electronic properties are considered material technology for innovations and advancements in electronics,[1] for instance, enabling thin-film electronics and circuits beyond the silicon technology.[2] Towards their scalable applications, solution-processing holds great promise. Solution-processing allows a low-cost, low-temperature production of mono- and few-layer 2D material nanoflakes in large quantities. Importantly, the solution-processed 2D materials can be readily adapted to the CMOS processes and emerging printing technologies for large-scale device fabrication.[3, 4] Though promising, the performance of the enabled devices is limited by the poor conduction of the underlying solution-processed 2D materials, as a result of the random, fragmented nature of the nanoflakes and the percolative conduction through the discrete nanoflake networks with significant inter-flake junction resistances.[4-6]

Despite the ongoing efforts to enlarge the lateral dimensions of the nanoflakes[7] and engineer the nanoflake alignments in the networks,[8] much work is still needed to overcome the poor conduction problem towards the practical applications of the solution-processed 2D materials. Encouragingly, given the atomic thickness and the quantum confinement effects, the electronic structures and properties of the solution-processed 2D materials may be effectively modulated with the ambient electric, optical, and even magnetic conditions, thereby facilitating the fabrication of functional and even high-performance devices.[1, 9] For example, the *quantum-* *confined Stark effect*, an effect that governs the electronic structures in nanostructured materials in local fields,[10] may be employed as an effective conduction modulation approach for the solution-processed 2D materials. Indeed, prior theoretical studies by density-functional theory (DFT) calculations have predicated electrostatic modulation of the electronic structures and properties of 2D materials with the Stark effect.[11, 12] State-of-the-art experimental studies on individual 2D material nanoflakes produced by mechanical exfoliation have also proved that the local fields exerted by, for instance, interfacial ferroelectrics (e.g. P(VDF-TrFE))[13-15] and ionic liquids (e.g. DEME-TFSI),[16] can lead to an effective modulation. However, few studies have been conducted on the solution-processed 2D material counterparts and, particularly, questions may arise as to whether the random, fragmented nature of the solution-processed nanoflakes and the percolative conduction through their discrete networks can allow an effective modulation by the Stark effect.

Herein, based on the prior studies, we adapt the Stark effect modulation approach to the solution-processed 2D materials. Using liquid-phase exfoliated molybdenum disulfide ($MoS_2$) as an example, we demonstrate nonlinear conduction modulation with a switching ratio of >$10^5$ by the local fields from the interfacial ferroelectric P(VDF-TrFE). Through DFT calculations and *in situ* Raman scattering and photoluminescence spectroscopic analysis, we understand that the charge in the solution-processed $MoS_2$ can redistribute in the local fields and as such, lead to the nonlinear conduction modulation. Beyond $MoS_2$, we show the modulation may be viable for the other solution-processed 2D materials and low-dimensional materials. This effective modulation will shed light on the understanding and engineering of the solution-processed 2D materials and, importantly, open their electronic device applications.

## 2. Results
### 2.1. Conduction modulation of solution-processed $MoS_2$

In this work, we study $MoS_2$ as an example as it is one of the most studied archetypes of 2D materials and, in its mono- and few-layer forms, allows a convenient tunning of the electronic structures and properties in the local fields.[17] Figure 1a shows the typical chemical structure of $MoS_2$ in the trigonal prismatic 2H phase. Figure 1b schematically illustrates a possible Stark effect progression in $MoS_2$, where the charge redistributes in the local electric field and forms electric dipoles.[12, 18] As discussed, the prior studies have proven the effectiveness of the Stark effect modulation on the mechanically exfoliated $MoS_2$ samples.[14, 15, 19] Here, we investigate adaptation of the Stark effect modulation approach to the solution-processed $MoS_2$.

Briefly, we prepare MoS$_2$ via liquid-phase exfoliation (see Methods); Fig. 1c. As shown in Fig. 1d (see also Fig. S1), the exfoliated MoS$_2$ is in a nanoflake form with a mean thickness of ~4 atomic layers. The microscopic study also proves that the exfoliated MoS$_2$ is in the 2H phase with a hexagonal chemical structure. The solution-processed MoS$_2$ is then deposited followed by ferroelectric P(VDF-TrFE) via inkjet printing to develop a vertical thin-film device where MoS$_2$ and P(VDF-TrFE) are sandwiched between electrodes; Fig. 1e (see also Fig. S2). Due to the interlayer wetting and diffusing, the exfoliated mono- and few-layer MoS$_2$ nanoflakes are embedded in P(VDF-TrFE) and form discrete networks. P(VDF-TrFE) is introduced to apply the local fields to induce the Stark effect when polarized.

As shown in Fig. 1f, the MoS$_2$-P(VDF-TrFE) device exhibits ferroelectric polarization when biased. This is ascribed to the ferroelectric polarization of P(VDF-TrFE); Fig. S3. Based on the prior studies, we expect the local fields from P(VDF-TrFE) to modulate the electronic property and as such, the conduction of the embedded solution-processed MoS$_2$. Indeed, as shown in Fig. 1g, the MoS$_2$-P(VDF-TrFE) device demonstrates a nonlinear current output with switching hysteresis. Notably the dynamical conductance ($G_d = \mathrm{d}I/\mathrm{d}V$) switches from ~2.26×10$^{-3}$ μS to ~319.60 μS at the compliance, with a switching ratio of >10$^5$ at the compliance. In contrast, as shown in Fig. 1h (see also Fig. S4), the devices developed with either MoS$_2$ or P(VDF-TrFE) exhibits limited or even no conduction modulation – the MoS$_2$ device shows a linear current output with poor conductance (~4.5×10$^{-3}$ μS), whereas the P(VDF-TrFE) device presents ~0.05 μS with a negligible switching (ratio <10) due to its polarization. This thus excludes the possibility that the significant nonlinear conduction modulation observed in the MoS$_2$-P(VDF-TrFE) device originates solely from neither MoS$_2$ nor P(VDF-TrFE) and, importantly, proves that the conduction modulation is due to the interplay between MoS$_2$ and the local fields from P(VDF-TrFE).

**2.2. Electronic structure evolution in MoS$_2$**
Based on the Stark effect theory, we presume the demonstrated conduction modulation arises from a charge redistribution in the solution-processed MoS$_2$ in the local fields from P(VDF-TrFE). To understand the origin and physics of the modulation, we carry out DFT calculations and *in situ* spectroscopic analysis. Here we first investigate the electronic structure evolution in MoS$_2$ via DFT calculations (see Methods).

As our solution-processed MoS$_2$ is an ensemble of mono- and few-layers with a mean thickness of ~4 atomic layers (Fig. S1), we consider mono- to penta-layers in our DFT calculations. Particularly, the calculations of quadri-layer MoS$_2$ are discussed in detail in Fig. 2. As presented in Fig. 2b, our results show that an external electric field can indeed progressively redistribute the charge along the planar surfaces of bi- and few-layer MoS$_2$. Such a charge redistribution can lead to the splitting and shifting of the band edges and thus, narrow the band gap; Fig. 2c,d (see also Fig. S5a). This demonstrates a Stark effect progressing in MoS$_2$. On the other hand, as shown in the last subplot in Fig. 2d, the density of states (DOS) of the band edges progressively propagate to lower energy levels in the electric field, leading to increased free carrier concentrations. However, note that mono-layer MoS$_2$ with the absence of an intralayer spacing fails to show this electronic structure evolution as an intralayer spacing is critical to accommodate the redistributed charges and, in turn, enhance the induced dipoles.[12, 18] Nevertheless, our DFT calculations show that bi- and few-layer solution-processed MoS$_2$ can undergo the Stark effect in local electric fields and as such, lead to a charge redistribution.

To understand how a charge redistribution can impact the conduction of the solution-processed MoS$_2$, we perform a qualitative analysis based on the fundamental model of conductivity $\sigma = n_i q(\mu_n + \mu_p)$[20] (see Methods). Again, we use quadri-layer MoS$_2$ for the analysis. As shown in Fig. 3a, the conduction and valence band edge curvatures of quadri-layer MoS$_2$ vary slightly as the band gap narrows in an increasing field, leading to nearly invariant electron and hole effective masses (Fig. 3b). This means a nearly invariant carrier mobility in MoS$_2$ (Fig. 3c), given $\mu \propto 1/|m^*|^2$,[20] where $\mu$ is the mobility and $m^*$ is the effective mass of the carriers. Meanwhile, as the band gap narrows linearly with the field (Fig. 2c, and Fig. S5a), the carrier concentration in MoS$_2$ increases exponentially (Fig. 3c), given $n_i \propto \exp(-E_g/2k_\text{B}T)$,[20] where $n_i$ is the carrier concentration, $E_g$ is the band gap, $k_\text{B}$ is the Boltzmann constant, and $T$ is the temperature. Although the carrier mobility in MoS$_2$ remains almost invariant, the exponentially increased carrier concentration can lead to a drastically enhanced electrical conductivity, based on the above conductivity model. On the other hand, assuming a constant inter-flake carrier hopping probability, the exponentially increased carrier concentration can further lead to a drastically enhanced inter-flake percolative conduction among the discrete nanoflake networks.[4] As such, the increased carrier concentration may enhance not only the conduction of the solution-processed MoS$_2$ nanoflakes but also the percolative conduction through the discrete networks.

Therefore, to conclude, our calculations suggest that the solution-processed $MoS_2$ can undergo a charge redistribution in local fields due to the Stark effect, and that the charge redistribution can lead to a drastically enhanced conduction, as we experimentally demonstrate in Fig. 1g.

### 2.3. *In situ* spectroscopic analyses

To further study and confirm the mechanism of the conduction modulation, we conduct *in situ* Raman scattering and photoluminescence spectroscopic characterizations (Fig. 4a) to probe the electronic structure evolution in the solution-processed $MoS_2$.

The *in situ* Raman scattering spectrum (Fig. 4b) as well as the spatial mapping (Fig. S6) of the solution-processed $MoS_2$-P(VDF-TrFE) device when biased exhibit intensity variations, with the Raman scattering peak locations remaining almost invariant. The invariance in the peak locations means the maintenance of the 2H phase in the solution-processed $MoS_2$, as the peak locations are governed by the vibrational energy of the normal modes that are related to the crystal structures.[21] The observed invariance in the peak locations thus excludes the possibility for a metallic phase transition of the solution-processed $MoS_2$ in the fields, even though this can lead to the demonstrated conduction modulation. On the other hand, as the Raman intensity of crystals at a specific vibrational mode $Q$ follows $I_{Raman} \propto \partial\alpha/\partial Q$, where α is the polarizability,[22] the observed Raman intensity variations can originate from a charge redistribution that can alter the polarizability of the solution-processed $MoS_2$. Therefore, the *in situ* Raman scattering characterizations suggest that the solution-processed $MoS_2$ undergoes a charge redistribution in the local fields from P(VDF-TrFE). Given that a charge redistribution can lead to conduction modulation based on our above discussion, the Raman scattering characterizations underpin the relation between the charge redistribution in $MoS_2$ and the conduction modulation as we experimentally demonstrate.

In addition to the intensity variations in Raman scattering, a charge redistribution can impact the electronic band structure. Here we carry out further *in situ* photoluminescence spectroscopic characterization to probe the charge redistribution in the solution-processed $MoS_2$ in the local fields from P(VDF-TrFE). As demonstrated in Fig. 4c (see also Fig. S7), the two characteristic photoluminescence peaks exhibit profound quenches and also slight redshifts. However, note that the peaks are much broader than those reported from materials prepared by other means such as mechanical exfoliation,[23] as a result of the fact that the solution-processed $MoS_2$ is an ensemble of mono- and few-layer nanoflakes. As the solution-processed nanoflakes are randomly deposited, we conduct a spatial mapping of the photoluminescence (Fig. 4d). As shown in Fig. 4e, f, the photoluminescence variations are universal and yet spatial dependent – for some of the studied sites the two peaks exhibit quenching and shifting, whereas for the other sites the variations are mild. Besides, interestingly, a new wave packet at ~850 nm can be observed at certain sites (Fig. S7 g, h), demonstrating a significant photoluminescence redshift. The redshifts prove a bandgap narrowing in the solution-processed $MoS_2$ in the local fields from P(VDF-TrFE). This confirms the Stark effect and, importantly, indicates a charge redistribution in the solution-processed $MoS_2$ due to the Stark effect. On the other hand, as presented in Fig. 2, as the induced dipoles are formed by the redistributed charges, the electrons and holes become delocalized and spatially separate across the $MoS_2$ nanoflakes in the fields, leading to a spatial separation of the highest occupied molecular orbitals (HOMO) and the lowest unoccupied molecular orbitals (LUMO) (Fig. S5c). The spatial separation can prevent the excited electron and hole pairs from recombining, leading to photoluminescence quenching,[24, 25] as observed in Fig. 4c, e, and f. The photoluminescence quenching, in turn, indicates a charge redistribution in the solution-processed $MoS_2$ due to the Stark effect. The photoluminescence shifting and quenching, therefore, again underpin the relation between the charge redistribution in $MoS_2$ and the conduction modulation as experimentally demonstrated.

Based on the spectroscopic analysis as well as the DFT calculations, we understand that our solution-processed $MoS_2$ undergoes a charge redistribution in the local fields from P(VDF-TrFE), and that the charge redistribution is the underlying mechanism of the conduction modulation.

## 3. Conclusion

In this work, we have reported conduction modulation of solution-processed $MoS_2$ via the Stark effect. We have established an understanding that local electric fields can be employed to lead to a charge redistribution in the solution-processed $MoS_2$ and as such, modulate the conduction with a high-order nonlinearity. Beyond $MoS_2$, we show this Stark effect modulation approach may be viable to the other types of 2D materials, such as molybdenum ditelluride and bismuth telluride, as well as low-dimensional materials, such as carbon nanotubes (Fig. S8). However, further investigations on these materials may be demanded to unveil the detailed modulation mechanisms, as the modulation may involve vibrational and electron-phonon coupling, and the ferroelectricity interplay between the materials and the interfacial ferroelectrics. Nevertheless, an effective conduction modulation can open the device applications of the solution-processed 2D materials.

## 4. Experimental Methods

*Solution-processing*: MoS$_2$ is prepared by liquid-phase exfoliation, following the method reported in Ref. [5]. Briefly, 200 mg bulk powder and 20 mL 1-methylpyrrolidin-2-one are mixed, sonicated for 48 h, and then centrifuged at 4,000 rpm for 30 min to acquire a supernatant. The supernatant is then solvent-exchanged via filtration into isopropyl alcohol/2-butanol (90/10 vol.%) to prepare the MoS$_2$ dispersion. P(VDF-TrFE) (70:30) is dissolved in N,N-dimethylformamide (DMF) via stirring to form a 2.5 wt.% dispersion. DMF is purchased from Alfa Aesar, and the others are from Sigma-Aldrich. All materials are used as received.

*Device fabrication and characterizations:* The MoS$_2$-P(VDF-TrFE) vertical nonlinear device is fabricated by depositing the above MoS$_2$ and P(VDF-TrFE) dispersions layer by layer via inkjet printing onto indium tin oxide (ITO) quartz substrate, and then deposited with a top electrode of gold via e-beam evaporation. The other MoS$_2$ and P(VDF-TrFE) vertical devices are also fabricated in a similar manner. The printer used is Fujifilm Dimatix Materials Printer DMP-2831. The current output is measured by Keithley 4200A-SCS. The ferroelectric hysteresis loop is measured by Radiant Technologies Multiferroic II Ferroelectric Tester. The polarization pattern images are measured by Brucker Dimension Icon atomic force microscope with a resonant-enhanced piezo-response force microscopy mode.

*In situ spectroscopic analyses: In situ* photoluminescence and Raman scattering measurements are performed at room temperature using 532 nm continuous laser excitation with a diffraction-limited excitation beam diameter of ~1 μm. The excitation light is filtered by a 532-nm-long pass filter (Semrock), and the photoluminescence and Raman signal light are separated by 300 g/mm and 1200 g/mm grating, respectively, and then received by liquid nitrogen cooled charge-coupled device. *In situ* photoluminescence and Raman measurements under bias are implemented by Keithley 2636B.

*DFT calculation:* The calculation is conducted using Vienna Ab-initio Simulation Package.[26] The interaction between the ionic core and valence electrons is described by projector augmented wave pseudopotential,[27] and the exchange-correlation (XC) energy between the electrons is described by the generalized gradient approximation functional in the Perdew-Burke-Ernzerhof scheme.[28] During calculation, the Mo (4*p*, 5*s*, 4*d*) states and S (3*s*, 3*p*) states are treated as valence states. The plane wave cut-off energy is set to be 500 eV. The interaction between the atomic layers of MoS$_2$ is corrected based on the DFT-D3 method[29] with Becke-Johnson damping function.[30] Besides, spin-orbit coupling effect is considered in the calculation. For the simulation of MoS$_2$, a slab geometry with 22 Å vacuum layer is chosen to prevent interaction between the material nanoflakes caused by the translational symmetry along the [001] direction. The structure of all the systems is fully optimized until the residual force on atoms is less than 0.01 eV/Å. For self-consistent iteration, the convergence tolerance between the successive steps is set to be 1×10$^{-5}$ eV. Monkhorst-Pack method is used to generate the *k*-point grid. For structural optimization, all material nanoflakes use a 14×14×1 grid. For self-consistent calculation, mono-layer and bi-layer MoS$_2$ use a 25×25×1 grid, while others use a 19×19×1 grid. For integrating the density of states, mono- and bi-layer MoS$_2$ use a 31×31×1 grid, while others use a 21×21×1 grid. In the calculation of the Stark effect, the electric field is applied by introducing a dipole sheet in the middle of the vacuum layer. Since the MoS$_2$-P(VDF-TrFE) device is fabricated in a vertical configuration, the interfacial ferroelectric fields always have components perpendicular to the solution-processed MoS$_2$ nanoflakes except for those minority nanoflakes with their atomic planes vertically aligned. The value setting of the electric fields to trigger the Stark effect may be larger than in previous reports,[18, 31] as our calculation considers the MoS$_2$ nanoflakes placed in vacuum regardless of the charge screening effects from the interfacial dielectrics. The visualization of the atomic structure, molecular orbitals and charge density are carried out via VESTA.[32]

*Qualitative model of conduction:* The fundamental electrical conductivity model is considered, expressed as $\sigma = n_i q(\mu_n + \mu_p)$.[20] Two key factors, i.e., the carrier concentration $n_i = n \approx p$ ($n$ and $p$ denotes the electron and hole concentration, respectively) and the carrier mobility $\mu$ ($\mu_n$ and $\mu_p$ denotes the electron and hole mobility, respectively), govern the conductivity (*10*). The effective mass of carriers (electrons and holes) when transporting in bands is $m^* = [(\partial^2 E(k)/\partial k^2)/\hbar^2]^{-1}$,[20] where $E(k)$ is the energy dispersion curves, i.e., the energy bands, and $\hbar$ is the reduced Planck constant. Based on the deformation theory,[33, 34] the carrier mobility of two-dimensional systems follows $\mu_{2D} = 2q\hbar^3 C/3k_B T|m^*|^2 E_1^2$, where $q$ is the elementary charge, $C$ is the elastic modulus, $k_B$ is the Boltzmann constant, $T$ is the temperature, and $E_1$ is the deformation potential constant. According to the effective density of states approximation,[20] the carrier concentration of intrinsic semiconductors follows $n_i = n \approx p = (N_C N_V)^{\frac{1}{2}} \exp(-E_g/2k_B T)$, where $N_C = m_e^* k_B T/\pi\hbar^2$ and $N_V = m_h^* k_B T/\pi\hbar^2$ is the effective density of states at the conduction band minimum and valence band maximum of two-dimensional systems, respectively. As the effective mass of electrons and holes varies slightly as the electric field increases, $N_C$ and $N_V$ remain nearly unchanged with the field.

## Acknowledgements

GHH acknowledges funding from CUHK (4055115), SHIAE (RNE-p3-21) and RGC (24200521), Yang Liu from SHIAE (RNE-p3-21), JFP and YYW from RGC (24200521), TM from PolyU (P0042991), Yue Lin from NSFC (52273029), Fujian Provincial Key Project of Science & Technology (2022H0037) and Fujian Science & Technology Innovation Laboratory for Optoelectronic Information of China (2021ZZ119), and GW from NSFC (12074033 and 11904019) and Beijing Natural Science Foundation (Z190006). The authors thank Shenzhen Cloud Computing Center, National Supercomputing Center in Shenzhen for providing the high-performance computing clusters service, and Miss Xiaolin Liu for discussions on the *ab initio* calculations.

## Conflict of Interest

The authors declare no conflict of interest.

## Data Availability Statement

The data that support the findings of this study are available from the corresponding authors upon reasonable request.

**Figures**

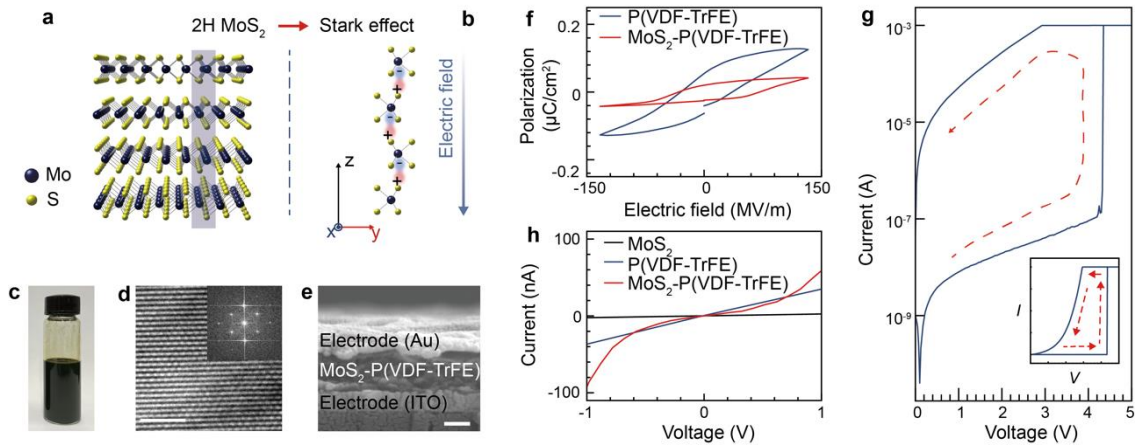

**Figure 1. Conduction modulation of solution-processed MoS₂.** a) Typical crystal structure of MoS$_2$ in 2H phase, and (b) the corresponding schematic of induced electric dipole generation in a external electric field. (c) Photograph of the MoS$_2$ dispersion. (d) Typical high-resolution transmission electron microscopic image of a MoS$_2$ nanoflake, with the hexagonal fast Fourier transform pattern shown in the inset, confirming the 2H phase. (e) Cross-sectional scanning electron microscopic image of a MoS$_2$-P(VDF-TrFE) device on ITO glass and with gold as the top electrode, showing MoS$_2$ and P(VDF-TrFE) inter-diffuse with no discernible interfaces. (f) Polarization hysteresis loops of the MoS$_2$-P(VDF-TrFE) and P(VDF-TrFE) devices. (g) Nonlinear hysteretic current output from the MoS$_2$-P(VDF-TrFE) device, showing a dynamical conductance ($G_d = \mathrm{d}I/\mathrm{d}V$) switching from ~2.26×10$^{-3}$ μS to ~319.60 μS with a switching ratio of >10$^5$ at the compliance of 1 mA. The inset is in linear scale. (h) Current output of MoS$_2$ (linear conductance ~4.5×10$^{-3}$ μS), P(VDF-TrFE) (negligible nonlinear conductance ~0.05 μS), and MoS$_2$-P(VDF-TrFE) devices.

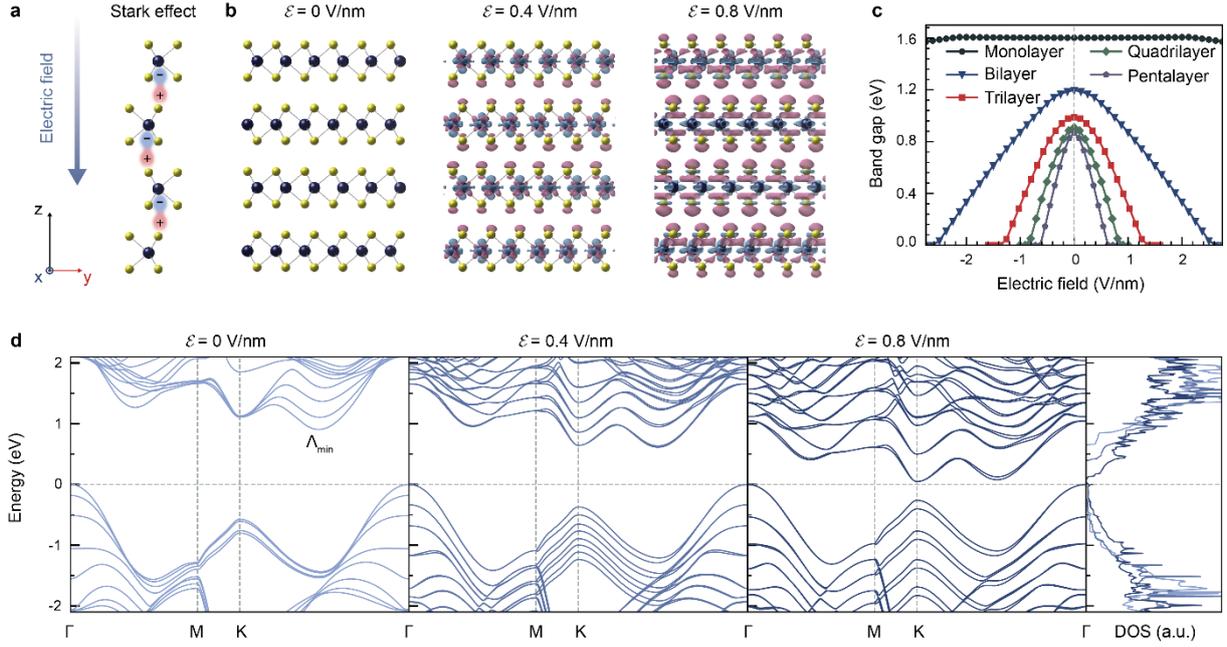

**Figure 2. Electronic structure evolution in MoS$_2$ due to the Stark effect.** (a) Replotted schematic induced electric dipole generation in MoS$_2$ a local electric field. (b) Differential charge density ($\Delta\rho$) of quadri-layer MoS$_2$ in an electric field ($\mathcal{E}$), calculated using $\Delta\rho(\mathcal{E}) = \rho(\mathcal{E}) - \rho(0)$, where $\rho(\mathcal{E})$ and $\rho(0)$ is the charge density in the $\mathcal{E}$ and zero electric field, respectively. The red and blue iso-surface is set to be plus and minus 5.4×10$^{-5}$ $e$/Å$^3$, respectively. The electric field is pointing downward. (c) Band gap evolution in MoS$_2$ with respect to the electric field, showing band gap narrowing due to the Stark effect. (d) Band structure evolution of quadri-layer MoS$_2$ in an electric field, with the last subplot showing the density of states (DOS) evolution. Electronic evolution in quadri-layer MoS$_2$ is presented as solution-processed MoS$_2$ has a mean thickness of 4 layers.

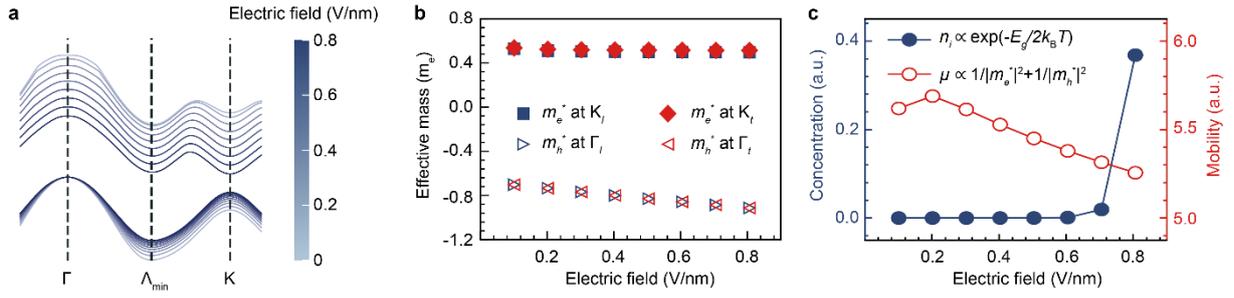

**Figure 3. Qualitative analysis on conduction modulation.** (a) Conduction and valence band edge evolution of quadri-layer MoS$_2$. The curvature of the energy band at the two valleys $\Lambda_{min}$ and K, as well as the two peaks $\Gamma$ and K, changes slightly as the electric field increases. (b) Effective mass of electrons moving at the valley K, and effective mass of holes moving at the peak $\Gamma$ at the different fields. The longitudinal direction (denoted with subscript $l$) is parallel to $\overrightarrow{\Gamma K}$ and the transverse direction (denoted with subscript $t$) is perpendicular to $\overrightarrow{\Gamma K}$. (c) Carrier concentration and mobility of quadri-layer MoS$_2$ with respect to the electric field, showing that the concentration increases exponentially, and the mobility remains almost invariant.

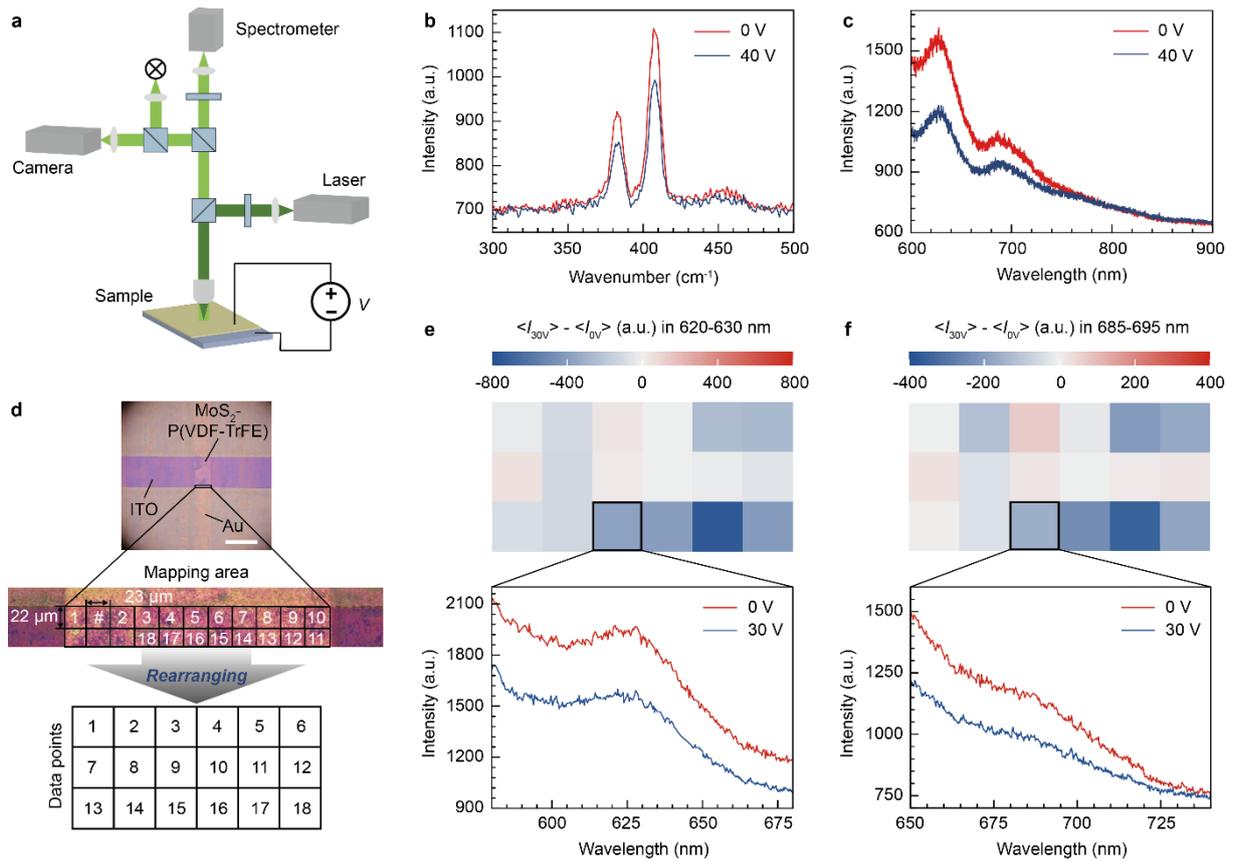

**Figure 4. *In situ* spectroscopic characterizations.** (a) Schematic Raman scattering and photoluminescence spectroscopic system. *In situ* Raman scattering (b) and photoluminescence (c) spectra of the $MoS_2$-P(VDF-TrFE) device with different bias. (d) Optical microscopic image of the $MoS_2$-P(VDF-TrFE) device, and the sites randomly selected for the *in situ* spectroscopic mapping, scale bar 1 mm. The site labelled # gives incomplete mapping results due to system errors. The mapping results from the other 18 sites are rearranged in display positions for the convenience of representing the results in (e) and (f). Mapping of the mean differential intensity in *in situ* photoluminescence of the $MoS_2$-P(VDF-TrFE) device at 620-630 nm (e) and 685-695 nm (f). The mean differential intensity is the difference between the mean photoluminescence intensity at a bias of 0 V and 30 V, respectively. Obvious photoluminescence quenching appears at site 2, 5, 6 and 15-18. The spectra at the two characteristic peaks of ~ 625 nm and ~ 690 nm of site 15 are presented.

Supporting Information for

**Conduction modulation of solution-processed two-dimensional materials**

*Songwei Liu, Xiaoyue Fan, Yingyi Wen, Pengyu Liu, Yang Liu, Jingfang Pei, Wenchen Yang, Lekai Song, Danmei Pan, Teng Ma, Yue Lin, Gang Wang\*, and Guohua Hu\**

E-mail: ghhu@ee.cuhk.edu.hk, gw@bit.edu.cn

This file contains:
Supplementary Figure S1-S8
References

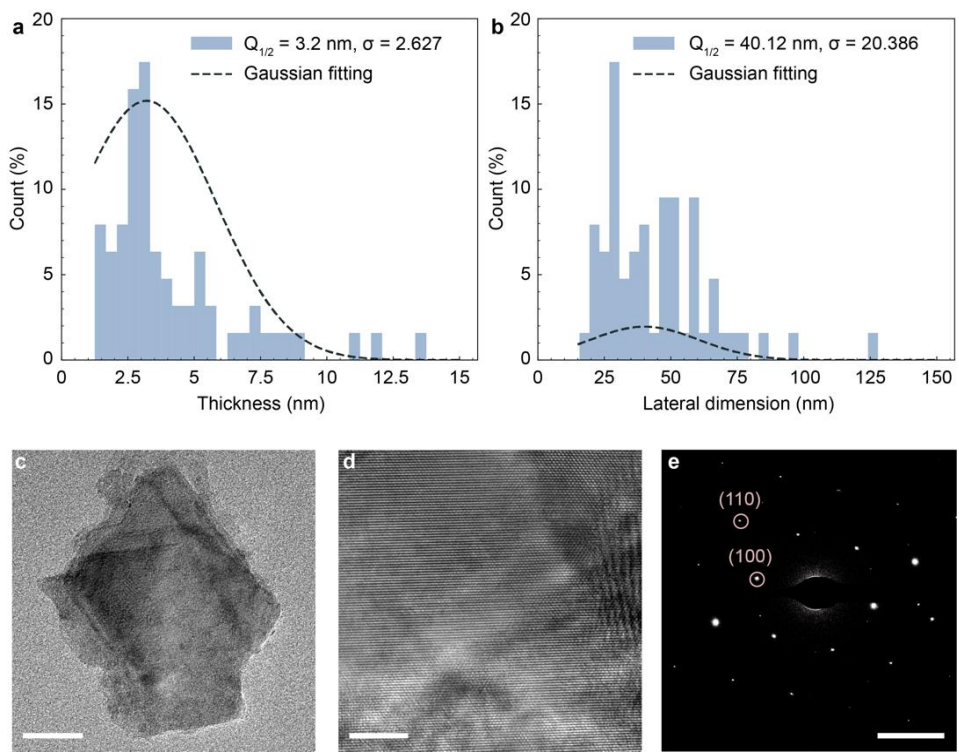

**Figure S1. Characterizations of the liquid-phase exfoliated MoS$_2$ flakes.** The thickness distribution (a) and the lateral dimension distribution (b) of the exfoliated MoS$_2$ flakes as measured by atomic force microscopy. The dashed lines are the Gaussian fittings. The mean thickness is 3.2 nm and the standard deviation is 2.627 nm. The mean lateral dimension is 40.12 nm and the standard deviation is 20.386 nm. According to prior studies,[1] the mean thickness shows the exfoliated MoS$_2$ flakes are ~ 4 atomic layers. A typical transmission electron microscopic (TEM) image (c), high-resolution TEM images (d), and selected-area electron diffraction image (e) of the MoS$_2$ flakes. Scale bars, 50 nm (c), 5nm (d), and 4 nm$^{-1}$ (e).

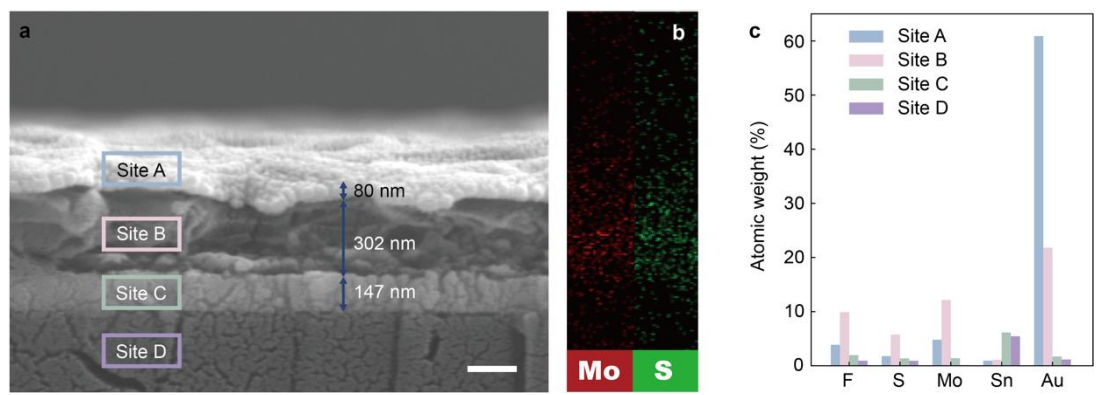

**Figure S2. Structural properties of the MoS$_2$-P(VDF-TrFE) device.** (a) Cross-sectional scanning electron microscopic image of the MoS$_2$-P(VDF-TrFE) device with gold and ITO electrodes, scale bar 200 nm. (b) Cross-sectional energy dispersive X-ray spectroscopic (EDS) mapping of the MoS$_2$-P(VDF-TrFE) device. The molybdenum and sulfur traces are sourced from MoS$_2$, showing that MoS$_2$ is distributed through P(VDF-TrFE). Mapping of fluorine sourced from P(VDF-TrFE) fails to be conducted as the mapping is easily disrupted by the ambient noises. (c) Point scanning EDS analysis of the MoS$_2$-P(VDF-TrFE) device. The scanning sites are indicated in (a). Site A shows a strong signal of gold from the gold electrode. Site C shows a strong signal of tin from the ITO electrode. Site B corresponds to the MoS$_2$-P(VDF-TrFE) device, and shows strong signals for fluorine, molybdenum and sulfur, confirming the presence of P(VDF-TrFE) and MoS$_2$ in site B. The EDS mapping in (b) shows that molybdenum and sulfur are distributed at the same locations spatially through the deposited MoS$_2$-P(VDF-TrFE) layer, suggesting MoS$_2$ and P(VDF-TrFE) form an inter-diffused layer where the mono- and few-layer MoS$_2$ nanoflakes are embedded in P(VDF-TrFE) and form networks.

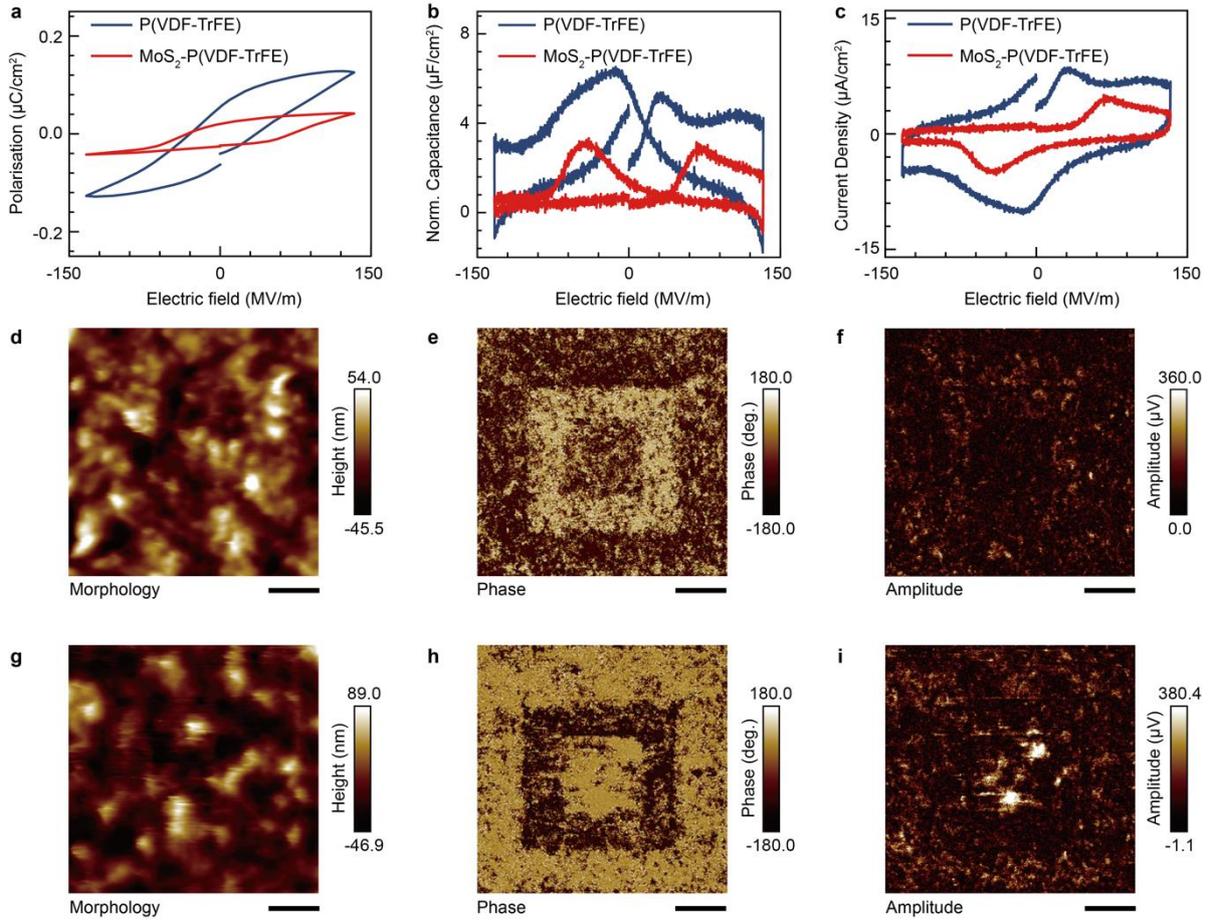

**Figure S3. Ferroelectricity characterizations of solution-processed MoS$_2$-P(VDF-TrFE) and P(VDF-TrFE).** Hysteresis loops of the polarization (a), normalized capacitance (b), and current density (c) versus the electric field of the solution-processed MoS$_2$-P(VDF-TrFE) and P(VDF-TrFE) devices. The morphology (d), piezo-response force microscopic (PFM) out-of-plane phase (e), and amplitude (f) images of a P(VDF-TrFE) thin-film, scale bars 1 μm. The morphology (g), PFM out-of-plane phase (h), and amplitude (i) images of a MoS$_2$-P(VDF-TrFE) thin-film, scale bars 1 μm. The polarization hysteresis loop (a) and the PFM out-of-plane phase images (f) and (i) demonstrate the ferroelectricity of P(VDF-TrFE). The polarization hysteresis loop (a) is presented in Fig. 1f and is replotted here for a convenient comparison.

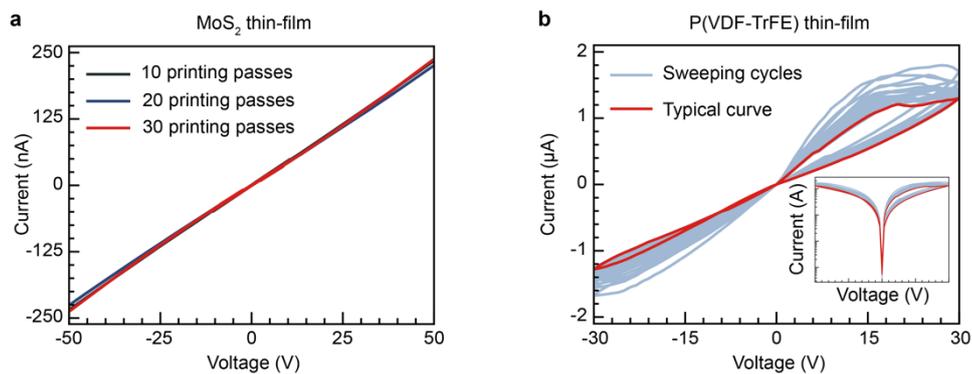

**Figure S4. Current output of the solution-processed MoS$_2$ and P(VDF-TrFE) devices.** Current output profiles of the solution-processed MoS$_2$ device, with an invariant, poor conductance of ~4.5×10$^{-3}$ μS (a), and the solution-processed P(VDF-TrFE) device, with a conductance of ~0.05 μS and a negligible conduction modulation ratio of < 10 (b).

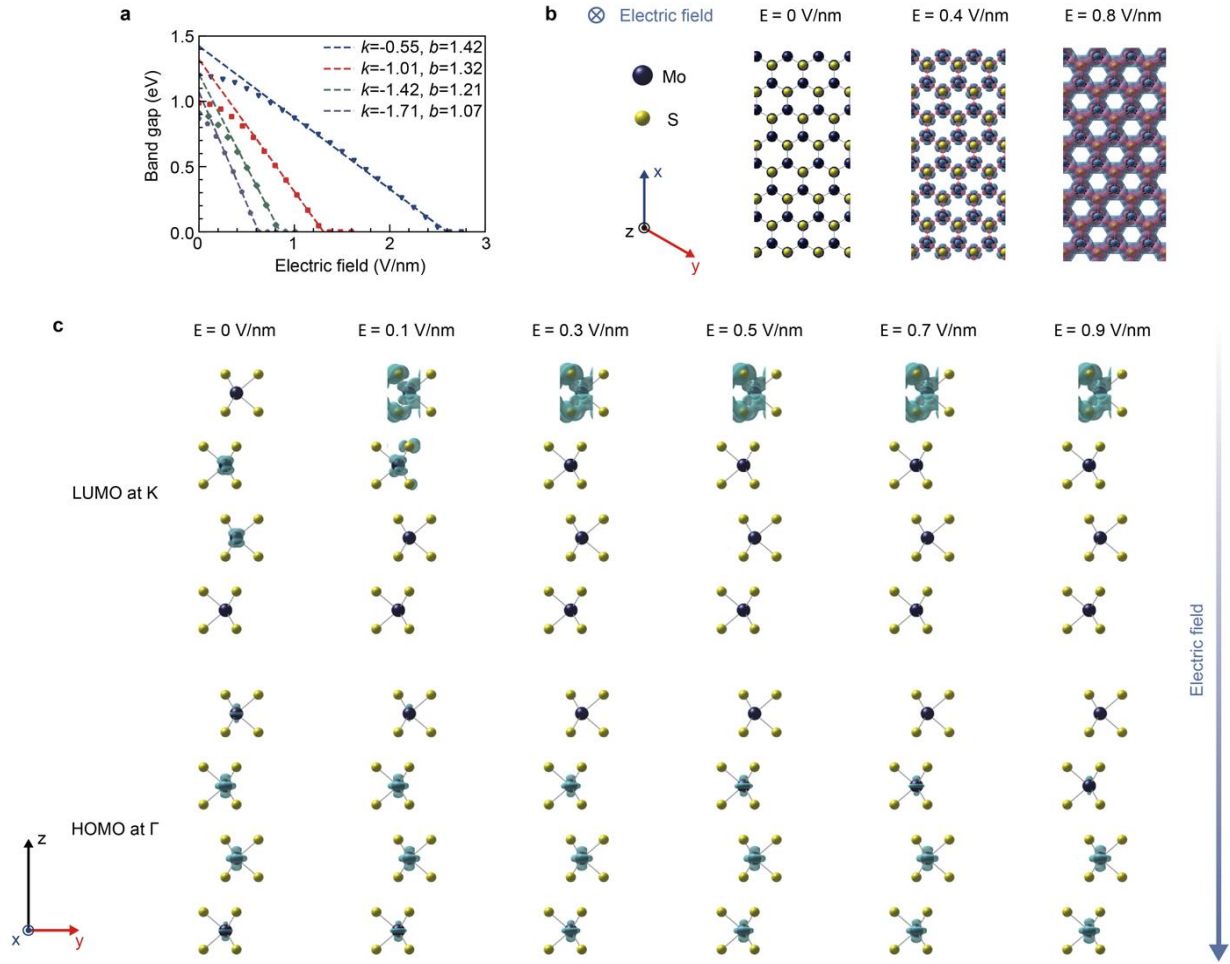

**Figure S5. The Stark effect.** (a) Linear fitting of the band gap evolution in mono- to penta-layer $MoS_2$ in an electric field, showing bi- to penta-layer $MoS_2$ exhibit band gap shrinking due to the Stark effect. The fitting follows $E_g = k\mathcal{E} + b$, where $k$ and $b$ is the slope and intercept of the fitting, respectively. The slope $k$ is the Stark coefficient, and is estimated as 0.55 nm, 1.01 nm, 1.42 nm and 1.71 nm for bi- to penta-layer $MoS_2$, respectively. The thickness-dependent Stark coefficient is consistent with prior studies.[2] (c) Top-view differential charge density development in quadri-layer $MoS_2$. The differential charge density $\Delta\rho$ at an electric field $\mathcal{E}$ is calculated by $\Delta\rho(\mathcal{E}) = \rho(\mathcal{E}) - \rho(0)$, where $\rho(\mathcal{E})$ and $\rho(0)$ is the charge density in the $\mathcal{E}$ and zero field, respectively. The red and blue iso-surface is set to be plus or minus $5.4 \times 10^{-5}$ $e/Å^3$, respectively. A charge layer is progressively developed as the electric field increases. (d) Evolution of LUMO (upper set) and HOMO (lower set) in quadri-layer $MoS_2$. At a zero electric field, the spatial distribution of the LUMO and HOMO is close. As the electric field increases, LUMO move against the field, while HOMO move along the field, leading to the spatial separation of molecular orbitals.

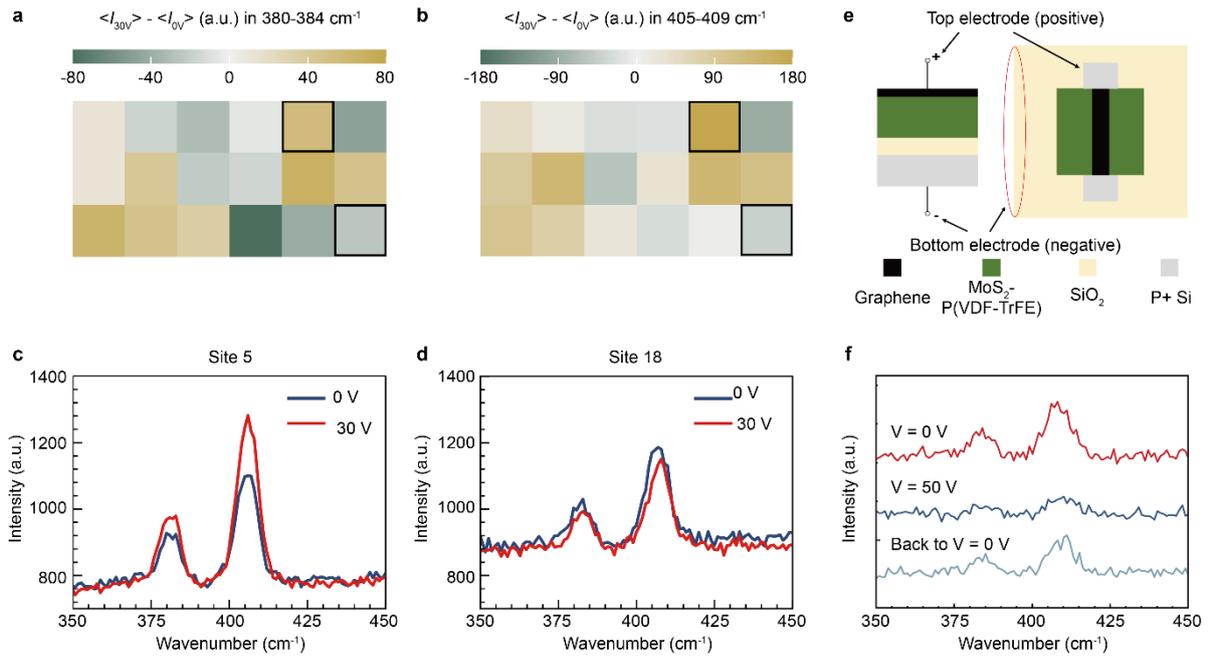

**Figure S6. *In situ* Raman scattering characterizations.** Differential intensity mapping of the *in situ* Raman scattering of the MoS$_2$-P(VDF-TrFE) device at 380-384 cm$^{-1}$ (a) and 405-409 cm$^{-1}$ (b). The differential intensity is the difference between the Raman scattering intensity at a bias of 0 V and 30 V, respectively. The mapping results are collected from the 18 sites as presented in Fig. 4d. When the bias is applied, the Raman scattering intensity exhibits enhancement or decline variations. The *in situ* Raman scattering spectra in site 5 (c) and site 18 (d) are two typical results showing Raman scattering enhancement and decline, respectively. According to prior studies,[3-6] this may arise from an electrostriction effect on the solution-processed MoS$_2$ flakes caused by the redistributed charge, and an internal stress caused by ripples as observed in the TEM images as shown in Fig. S2. To exclude the possibility that the Raman scattering evolution originates from charge injection from the electrodes and/or an electrical breakdown, a capacitor in P$^+$-Si/SiO$_2$/MoS$_2$-P(VDF-TrFE)/inkjet-printed translucent graphene electrode configuration (e) is fabricated and is characterized with *in situ* Raman scattering (f). The intensity of the two characteristic peaks is weakened as the bias is applied and partially recovers as the bias is withdrew, demonstrating that the Raman intensity evolution originates from the effect exerted by the interfacial ferroelectric field from the polarized P(VDF-TrFE).

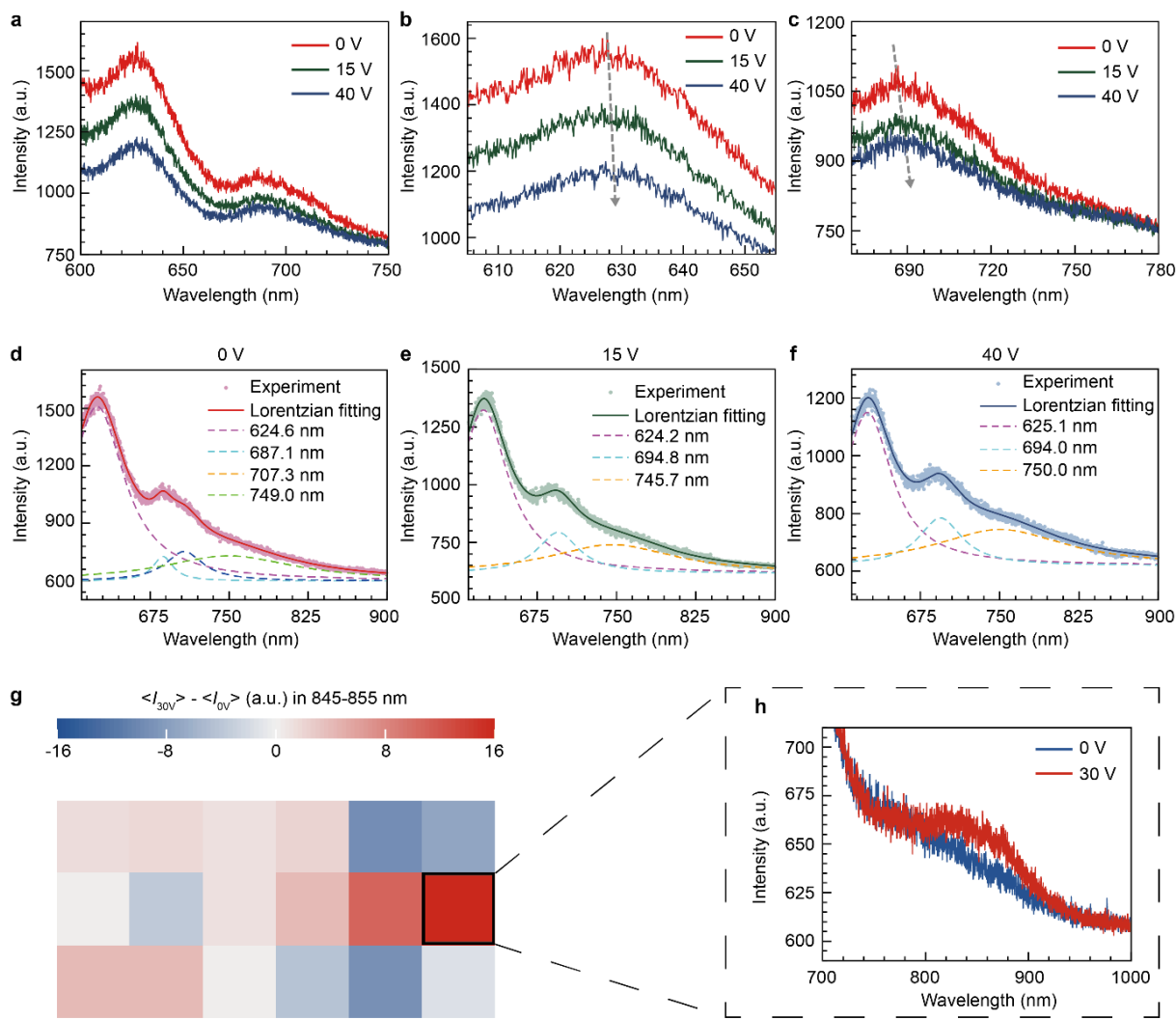

**Figure S7. *In situ* photoluminescence characterizations.** *In situ* photoluminescence spectra of the MoS$_2$-P(VDF-TrFE) device (a), with the zoomed-in characteristic peak evolution at ~ 625 nm (b) and ~ 690 (c), showing photoluminescence quenching and red-shifting. Photoluminescence fitting at a bias of 0 V (d), 15 V (e) and 40 V (f). The fitting is carried out with Lorentzian functions, given the photoluminescence is caused by radiative carrier recombination.[7] The fitting shows the photoluminescence is red-shifting collectively in response to the bias. The peak at ~ 625 nm shifts slightly, while the peak at ~ 690 nm exhibit more profound shifting. As the monolayer MoS$_2$ flakes have a direct band gap,[8] they can dominate the photoluminescence from the solution-processed MoS$_2$, leading to the slight shifting photoluminescence at ~ 625 nm. (g) Differential intensity mapping between the photoluminescence at 0V and 30V bias at 845-855 nm. Obvious photoluminescence wave pack appears at sites 10-14. Detailed wave pack spectra of site 12 as circled are presented in (h). The wave pack shows significant photoluminescence red shifting.

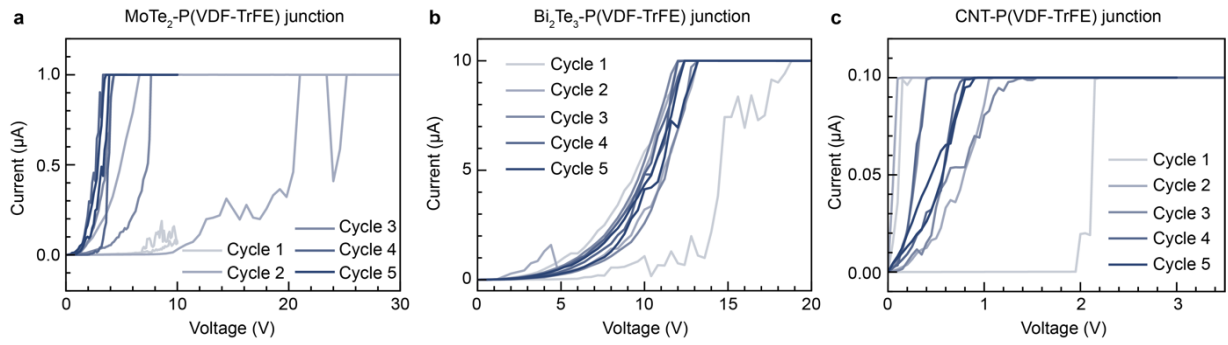

**Figure S8. Current output of devices developed with the other nanomaterials.** Current output of the (a) MoTe$_2$-, (b) Bi$_2$Te$_3$- and (c) carbon nanotube (CNT)-P(VDF-TrFE) devices.